\documentclass[onecolumn,showpacs,preprintnumbers,amsmath,amssymb]{revtex4}

\usepackage{graphicx}
\usepackage{dcolumn}
\usepackage{bm}
\usepackage{epsfig}

\begin{document}

\preprint{APS/123-QED}

\title{Horizontal visibility graphs: exact results for random time series}

\author{B. Luque$^1$, L. Lacasa$^1$, F. Ballesteros$^2$, and J. Luque$^3$}
\email{lucas@dmae.upm.es} \affiliation{$^1$Dpto. Matem\'{a}tica
Aplicada y Estad\'{i}stica, ETSI Aeron\'{a}uticos,\\Universidad
Polit\'{e}cnica de Madrid,
Spain\\
$^2$Observatorio Astron\'{o}mico, Universidad de Valencia, Spain\\
$^3$Dept de Teoria del Senyal i Comunicacions,\\Universitat
Polit$\grave{e}$cnica de Catalunya, Spain}%

\date{\today}

\begin{abstract}
The \emph{visibility algorithm} has been recently introduced as a
mapping between time series and complex networks. This procedure
allows to apply methods of complex network theory for
characterizing time series. In this work we present the
\emph{horizontal visibility algorithm}, a geometrically simpler
and analytically solvable version of our former algorithm,
focusing on the mapping of random series (series of independent
identically distributed random variables). After presenting some
properties of the algorithm, we present exact results on the
topological properties of graphs associated to random series,
namely the degree distribution, clustering coefficient, and mean
path length. We show that the horizontal visibility algorithm
stands as a simple method to discriminate randomness in time
series, since any random series maps to a graph with an
exponential degree distribution of the shape
$P(k)=(1/3)(2/3)^{k-2}$, independently of the probability
distribution from which the series was generated. Accordingly,
visibility graphs with other $P(k)$ are related to non-random
series. Numerical simulations confirm the accuracy of the theorems
for finite series. In a second part, we show that the method is
able to distinguish chaotic series from i.i.d. theory, studying
the following situations: (i) noise-free low-dimensional chaotic
series, (ii) low-dimensional noisy chaotic series, even in the
presence of large amounts of noise, and (iii) high-dimensional
chaotic series (coupled map lattice), without needs for additional
techniques such as surrogate data or noise reduction methods.
Finally, heuristic arguments are given to explain the topological
properties of chaotic series and several sequences which are
conjectured to be random are analyzed.
\end{abstract}

\pacs{05.45.Tp, 89.75.Hc, 05.45.-a}
\maketitle

\section{Introduction}
Recently, the visibility algorithm, a new tool for time series
analysis, has been introduced \cite{visibilidad_pnas}. The method,
inspired in the concept of visibility \cite{visibility}, proceeds
by mapping time series into graphs according to a specific
geometric criterion, in order to make use of complex networks
techniques \cite{redes2, redes3, redes4, redes5} for characterize
time series (some other works based on a similar philosophy can be
found in \cite{small1, small2}). In short, a visibility graph is
obtained from the mapping of a time series into a network
according to the following visibility criterion: Two arbitrary
data $(t_a, y_a)$ and $(t_b, y_b)$ in the time series have
visibility, and consequently become two nodes in the associated
graph, if any other data $(t_c, y_c)$ such that $t_a <t_c <t_b$
fulfills
\begin{equation}
y_c < y_a + (y_b-y_a)\frac{t_c-t_a}{t_b-t_a}.\label{eq1}
\end{equation}
It has been shown \cite{visibilidad_pnas} that time series
structure is inherited in the associated graph, such that
periodic, random and fractal series map into motif-like, random
exponential and scale-free networks \cite{redes1, redes2.0,
redes2.1}, respectively. These findings suggest that the
visibility graph may capture the dynamical fingerprints of the
process that generated the series. Furthermore, it has been
recently pointed out that this algorithm stands as a method for
estimating the Hurst exponent $H$ in fractional Brownian series
\cite{fbm_mandel}, since a linear relation between $H$ and the
exponent $\gamma$ of the power law degree distribution in the
scale free associated visibility graph exists \cite{hurst}. While
being relatively new, some applications of the method to analyze
time series, in different contexts from fluid dynamics
\cite{turbulence} or
atmospheric sciences \cite{hurricanes} to finance \cite{stock market}, have been presented so far.\\
\begin{figure}[h]
\centering
\includegraphics[width=0.50\textwidth]{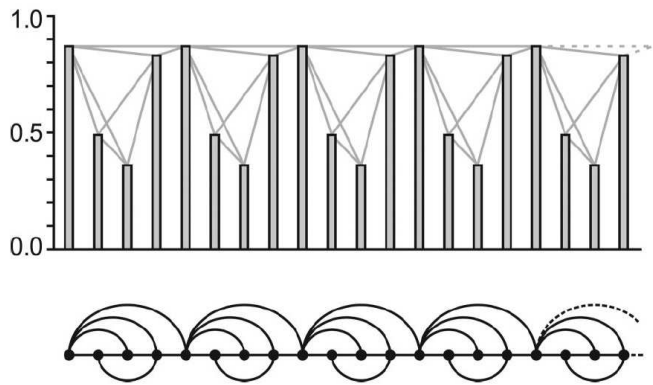}
\caption{Illustrative example of the visibility algorithm. In the
upper part we plot a periodic time series and in the bottom part
we represent the graph generated through the visibility algorithm.
Each datum in the series corresponds to a node in the graph, such
that two nodes are connected if their corresponding data heights
fulfill the visibility criterion of equation \ref{eq1}. Note that
the degree distribution of the visibility graph is composed by a
finite number of peaks, much in the vein of the Discrete Fourier
Transform of a periodic signal. We can thus interpret the
visibility algorithm as a geometric transform.}\label{periodic}
\end{figure}

\noindent What does the visibility algorithm stand for? In order
to deepen on the geometric interpretation of the visibility graph,
let us focus on a periodic series. It is straightforward that its
visibility graph is a concatenation of a motif: a repetition of a
pattern (see figure \ref{periodic}). Now, which is the degree
distribution $P(k)$ of this visibility graph? Since the graph is
just a motif's repetition, the degree distribution will be formed
by a finite number of non-null values, this number being related
to the period of the associated periodic series. This behavior
reminds us the Discrete Fourier Transform (DFT), which for
periodic series is formed by a finite number of peaks (vibration
modes) related to the series period. Using this analogy, we can
understand the visibility algorithm as a geometric (rather than
integral) transform. Whereas a DFT decomposes a signal in a sum of
(eventually infinite) modes, the visibility algorithm decomposes a
signal in a concatenation of graph's motifs, and the degree
distribution simply makes a histogram of such 'geometric modes'.
While the time series is defined in the time domain and the DFT is
defined on the frequency domain, the visibility graph is defined
on the 'visibility domain'. This is, of course, a hand-waving
analogy and further work should study its extent rigorously. For
instance, this transform is not, as presented, a reversible one.
Reversibility can however be easily obtained weighting the links
in the visibility graph with the slope of the visibility line that
links the associated data heights. The weighted version of the
algorithm and its geometric transform nature will be addressed
elsewhere. At this point we can comment that whereas a generic DFT
fails to capture the presence of nonlinear correlations in time
series (such as the presence of chaotic behavior), in the second
part of this paper we will show that the visibility algorithm can
clearly distinguish between white noise (i.e. a sequence of
identically
independent random variables) and chaotic series.\\
Of course the latter analogy is, so far, a simple metaphor to help
our intuition. Indeed, while some analytical results have already
been put forward within the visibility algorithm
\cite{visibilidad_pnas, hurst} (typically making use of concepts
borrowed from extreme value theory), no rigorous theory for the
visibility algorithms exists so far. Our goal in this work is to
make the first steps in that direction, providing results on the
properties of the visibility graphs associated to random series.
In order to derive exact results, we present here a slight
modification of the algorithm that we call the \emph{horizontal}
visibility algorithm, which is essentially similar to the former
yet having a geometrically simpler visibility criterion. According
to this latter criterion, the generated horizontal visibility
graph stands as a subgraph of the visibility graph. We will prove
that, surprisingly, the horizontal visibility graph associated to
any random series is a Small-World \cite{redes2.0} random graph
with a universal exponential degree distribution of the form
$P(k)=(1/3)\cdot(2/3)^{k-2}$, independently of the probability
distribution $f(x)$ from which the series was generated.
Accordingly, the horizontal visibility algorithm stands as an
extremely simple test for randomness, that for instance, can
easily distinguish random series from chaotic ones. The remaining
of this paper goes as follows: in section II we introduce the
horizontal visibility algorithm, a geometrically simpler version
of the visibility algorithm that allows analytical tractability,
along with some of its properties. In section III we derive exact
results for the degree distribution $P(k)$ of the associated graph
to a generic random time series. In section IV exact results on
other properties of horizontal visibility graphs are also derived,
concretely (i) $P(k|x)$, the probability that a node associated to
a datum of height $x$ has degree $k$, (ii) the clustering
distribution $P(C)$, (iii) the probability of long distance
visibility $P(n)$ and (iv) an estimation of its mean path length
$L(N)$ (sections III and IV contain technical proofs that the
non-interested reader can eventually skip). In section V we study
the reliability of the algorithm to discriminate chaotic series
from our theory. For this task we calculate the degree
distribution of visibility graphs associated to (i)
low-dimensional chaotic series (logistic map, H\'{e}non map), (ii)
noisy low-dimensional chaotic series (with amounts of noise of
$100\%$ by amplitude), and (iii) high-dimensional chaotic series
(coupled map lattice). In every case, the algorithm easily
distinguishes those series from a series of i.i.d. random
variables (white noise). At this point we also conjecture that the
topological properties of graphs associated to chaotic series are
related to the statistics of Poincar\'{e} recurrence times
\cite{PRT5}. In section VI we make use of the method as a
randomness test, and study some number theoretical sequences that
are conjectured to be normal (decimal expansion of normal numbers
\cite{normal}). We finally provide some concluding remarks in
section VII.

\section{Horizontal visibility algorithm}
\begin{figure}[h]
\centering
\includegraphics[width=0.50\textwidth]{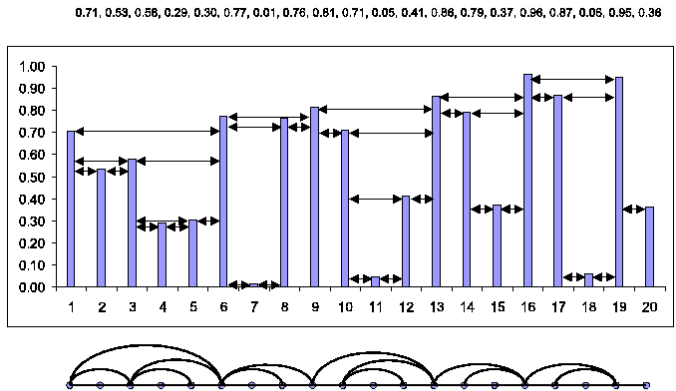}
\caption{Illustrative example of the horizontal visibility
algorithm. In the upper part we plot a time series and in the
bottom part we represent the graph generated through the
horizontal visibility algorithm. Each datum in the series
corresponds to a node in the graph, such that two nodes are
connected if their corresponding data heights are larger than all
the data heights between them. The data values (heights) are made
explicit in the top.}\label{visi_horizont}
\end{figure}
\noindent The horizontal visibility algorithm maps time series
into graphs and it is defined as follows. Let $\{x_i\}_{i=1..N}$
be a time series of $N$ data. The algorithm assigns each datum of
the series to a node in the horizontal visibility graph (graph
from now on). Two nodes $i$ and $j$ in the graph are connected if
one can draw a horizontal line in the time series joining $x_i$
and $x_j$ that does not intersect any intermediate data height
(see figure \ref{visi_horizont} for a graphical illustration).
Hence, $i$ and $j$ are two connected nodes if the following
geometrical criterion is fulfilled within the time series:
\begin{equation}
x_i, x_j > x_n \ \ \textrm{for all} \ \  n \ \ \textrm{such that}
\ \   i<n<j\ . \label{criterium}
\end{equation}
This algorithm is a simplification of the so called visibility
algorithm \cite{visibilidad_pnas} that has been recently
introduced. As a matter of fact, notice that given a time series,
its horizontal visibility graph is always a subgraph of its
associated visibility graph. Accordingly, as in the former case,
the horizontal visibility graph associated to a time series is always:\\
\noindent(\emph{i}) Connected: each node sees at least its nearest
neighbors (left-hand side
and right-hand side).\\
(\emph{ii}) Invariant under affine transformations of the series
data: the visibility criterium is invariant under rescaling of
both horizontal and vertical axis, as well as under horizontal and
vertical translations.\\
Some other properties can be stated, namely:\\
(\emph{iii}) Reversible/Irreversible character of the mapping:
note that some information regarding the time series is inevitably
lost in the mapping from the fact that the network structure is
completely determined in the (binary) adjacency matrix. For
instance, two periodic series with the same period as $T1 = {. . .
, 3, 1, 3, 1, . . .}$ and $T2 = {. . . , 3, 2, 3, 2, . . .}$ would
have the same visibility graph, albeit being quantitatively
different. Although the spirit of the visibility graph is to focus
on time series structural properties (periodicity, fractality,
etc.), the method can be trivially generalized by making use of
weighted networks (where the adjacency matrix is not binary and
the weights determine the height difference of the associated
data), if we eventually need to quantitatively distinguish time
series like T1 and T2, for instance. Using weighted networks, the
algorithm trivially converts to a reversible one.\\
(\emph{iv}) Undirected/directed character of the mapping: Although
this algorithm generates undirected graphs, note that one could
also extract a directed graph (related to the temporal axis
direction) in such a way that for a given node one should
distinguish two different degrees: an ingoing degree $k_{in}$,
related to how many nodes see a given node i, and an outgoing
degree $k_{out}$, that is the number nodes that node i sees. In
that situation, if the direct visibility graph extracted from a
given time series is not invariant under time reversion (that is,
if $P(k_{in})\neq P(k_{out}))$, one could assert that the process
that generated the series is not conservative. In a first
approximation we have studied the undirected version, and the
directed one will be eventually addressed in further work. While
the undirected choice seems to violate causality, note that the
same 'causality violation' is likely to take place when
performing the DFT of a time series, for instance.\\
 (\emph{vi}) Comparison between geometric criteria:
Note that the geometric criterion defined for the horizontal
visibility algorithm is more 'visibility restrictive' than its
analogous for the general case. That is to say, the nodes within
the horizontal visibility graph will have 'less visibility' than
their counterparts within the visibility graph. While this fact
does not have an impact on the qualitative features of the graphs,
quantitatively speaking, horizontal visibility graphs will have
typically 'less statistics'. For instance, it has been shown that
the degree distribution $P(k)$ of the visibility graph associated
to a fractal series is a power law $P(k)\sim k^{-\gamma}$, such
that the Hurst exponent $H$ of the series is linearly related to
$\gamma$ \cite{hurst}. Now, for practical purposes it is more
recommendable to make use of the visibility algorithm (in
detriment of the horizontal version) when measuring the Hurst
exponent of a fractal series, since a good estimation of $\gamma$
requires at least two decades of statistics in $P(k)$, something
which is more likely within the visibility algorithm. In what
follows we will show that the simplicity of the horizontal version
of the algorithm -which is computationally faster than the
original- allows analytical tractability, and nonetheless, this
latter method is well fitted to distinguish different degrees of
chaos from a sequence of uncorrelated random variables.

\section{Degree distribution of the visibility graph associated to a random time series}

\begin{figure}[h]
\centering
\includegraphics[width=0.50\textwidth]{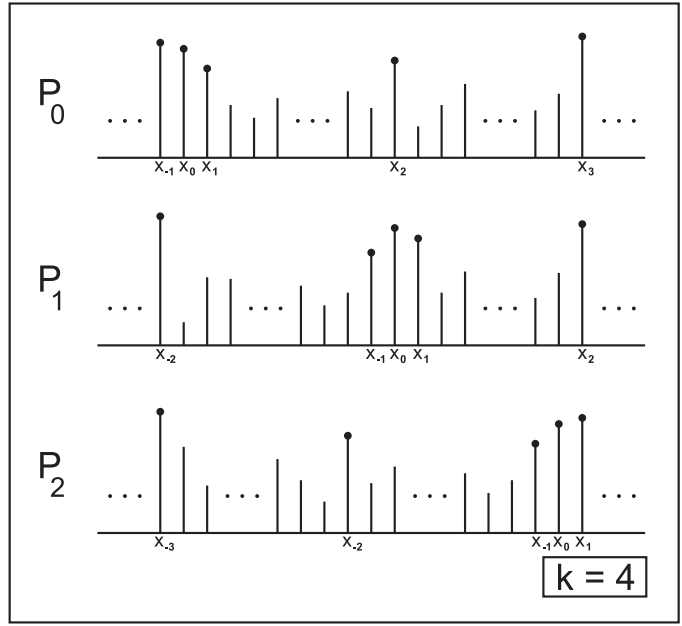}
\caption{Set of possible configurations for a seed data $x_0$ with
$k=4$. Observe that the sign of the subindex in $x_i$ indicates if
the data is located whether at left-hand side of $x_0$ (sign
minus) or at right-hand side. Accordingly, the bounding's data
subindex directly indicates the amount of data located in that
side. For instance, $C_0$ is the configuration where none of the
$k-2=2$ inner data are located in the left-hand side of $x_0$, and
therefore the left bounding data is labeled as $x_{-1}$ and the
right bounding data is labeled as $x_3$. $C_1$ is the
configuration for which an inner data is located in the left-hand
side of $x_0$ and another inner data is located in its right-hand
side. Finally, $C_2$ is the configuration for which both inner
data are located in the left-hand side of the seed. Notice that an
arbitrary number of hidden data can be eventually located among
the inner data, what is schematically represented in the figure as
a row of vertical lines. }\label{k4}
\end{figure}

Consider a bi-infinite time series created from a random variable
$X$ with probability distribution $f(x)$ with $x \in [0,1]$ and
let us construct its associated horizontal visibility graph (note
that if the distribution's support is a generic interval $x \in
[a,b]$, we can rescale to $[0,1]$ without loss of generality since
the associated graph remains invariant, and this also applies to
unbounded supports). For convenience, we will label a generic
datum $x_0$ as the `seed' datum from now on. In order to derive
the degree distribution $P(k)$ \cite{redes1} of the associated
graph, we need to calculate the probability that an arbitrary
datum with value $x_0$ has visibility of exactly $k$ other data.
If $x_0$ has visibility of $k$ data, there always will exist two
`bounding data', one on the right-hand side of $x_0$ and another
one on its left-hand side, such that the $k-2$ remaining visible
data will be located inside that window (in fact, $k=2$ is the
minimum possible degree). As these `inner' data should appear
sorted by size, there are exactly $k-1$ different possible
configurations $\{C_i\}_{i=0..k-2}$, where the index $i$
determines the number of inner data on the left-hand side of $x_0$
(see figure \ref{k4} for an illustration of the possible
configurations and a labelling recipe of the data in the case
$k=4$). Accordingly, $C_i$ corresponds to the configuration for
which $i$ inner data are placed at the left-hand side of $x_0$,
and $k-2-i$ inner data are placed at its right-hand side. Each of
these possible configurations have an associated probability
$p_i\equiv p(C_i)$ that will contribute to $P(k)$ such that
\begin{equation}
P(k)=\sum_{i=0}^{k-2}p_i \label{P_i}.
\end{equation}

\noindent Before trying to find a general relation for $P(k)$ and
for illustrative purposes, let us study some particular cases. The
first and simplest case is $P(k=2)$, that is, the probability that
the seed data has two and only two visible data, the minimum
degree. These obviously will be the bounding data, that we will
label $x_{-1}$ and $x_1$ for left-hand side and right-hand side of
the seed respectively. The probability that $x_0$ sees $k\geq2$ is
$1$ by construction, since the horizontal visibility algorithm
assures that any data will always have visibility of its first
neighbors. Now, in order to assure that $k=2$, we have to impose
that the bounding data neighbors have a larger height than the
seed, that is, $x_{-1}\geq x_0$ and $x_1\geq x_0$. Then,
\begin{equation}
P(k=2)=\textrm{Prob}(x_{-1},x_1\geq0)
=\int_0^1f(x_0)dx_0\int_{x_0}^1f(x_1)dx_1\int_{x_0}^1f(x_{-1})dx_{-1}.
\label{pk2}
\end{equation}
Now, the cumulative probability distribution function $F(x)$ of
any probability distribution $f(x)$ is defined as
\begin{equation}
F(x)=\int_0^xf(x')dx'\label{cumulative},
\end{equation}
where $dF(x)/dx=f(x)$, $F(0)=0$ and $F(1)=1$. In particular, the
following relation between $f$ and $F$ holds:
\begin{equation}
f(x)F^{n-1}(x)=\frac{1}{n}\frac{dF^n(x)}{dx}.\label{f-F}
\end{equation}
We can accordingly rewrite and compute equation \ref{pk2} as
\begin{equation}
P(k=2)=\int_0^1 f(x_0)[1-F(x_0)]^2dx_0=\frac{1}{3},
\end{equation}
independently of the shape of the probability distribution $f(x)$.\\

\noindent Let us proceed by tackling the case $P(k=3)$, that is,
the probability that the seed has three and only three visible
data. Two different configurations arise: $C_0$, in which $x_0$
has 2 bounding visible data ($x_{-1}$ and $x_2$ respectively) and
a right-hand side inner data ($x_1$), and the same for $C_1$ but
with the inner data being place at the left-hand side of the seed,
so
$$P(k=3)=p(C_0) + p(C_1)\equiv p_0+p_1.$$
\noindent Notice at this point that an arbitrary number $r$ of
hidden data $n_1,n_2...n_r$ can eventually be located  between the
inner data and the bounding data, and this fact needs to be taken
into account in the probability calculation. The geometrical
restrictions for the $n_j$ hidden data are $n_j<x_1, \ j=1,...,r$
for $C_0$ and $m_j<x_{-1}, \ j=1,...,s$ for $C_1$. Then,
\begin{eqnarray}
p_0 &=& \textrm{Prob}\bigg((x_{-1},x_2\geq x_0)\cap (x_1<x_0)\cap (\{n_j<x_1\}_{j=1,\dots,r})\bigg), \nonumber \\
p_1 &=& \textrm{Prob}\bigg((x_{-2},x_1\geq x_0)\cap
(x_{-1}<x_0)\cap(\{m_j<x_{-1}\}_{j=1,\dots,s})\bigg).
\label{P_1_k3bis}
\end{eqnarray}
Now, we need to consider every possible hidden data configuration
($C_0$ without hidden data, $C_0$ with a single hidden data, $C_0$
with two hidden data, and so on, and the same for $C_1$). With a
little calculus we come to
\begin{eqnarray}
&&p_0= \int_0^1f(x_0)dx_0 \int_{x_0}^1f(x_{-1}) dx_{-1}
\int_{x_0}^1 f(x_2)dx_2
\int_0^{x_0}f(x_1)dx_1 + \nonumber\\
&&\sum_{r=1}^\infty \int_0^1f(x_0)dx_0 \int_{x_0}^1
f(x_{-1})dx_{-1} \int_{x_0}^1 f(x_2)dx_2
\int_0^{x_0}f(x_1)dx_1\prod_{j=1}^r\int_0^{x_1}f(n_j)dn_j\nonumber\label{P0k3}
\end{eqnarray}
where the first term corresponds the contribution of a
configuration with no hidden data and the second sums up the
contributions of $r$ hidden data. Making use of the properties of
the cumulative distribution $F(x)$ we arrive to
\begin{equation}
p_0=\int_0^1f(x_0)dx_0 \int_{x_0}^1f(x_{-1}) dx_{-1} \int_{x_0}^1
f(x_2)dx_2 \int_0^{x_0}\frac{f(x_1)}{1-F(x_1)}dx_1,
\end{equation}
where we also have made use of the sum of a geometric series. We
can find an identical result for $p_1$, since the last integral on
equation \ref{P0k3} only depends on $x_0$ and consequently the
configuration provided by $C_1$ is symmetrical to the one provided
by $C_0$. We finally have
\begin{equation}
P(k=3)=2p_0=-2\int_0^1f(x_0)(1-F(x_0))^2\ln
(1-F(x_0))dx_0=\frac{2}{9},
\end{equation}
where the last calculation also involves the change of variable
$z=1-F(x)$. Again, the result is independent of $f(x)$.\\

\noindent Hitherto, we can deduce that a given configuration $C_i$
contributes to $P(k)$ with a product of integrals according to the
following rules:
\begin{itemize}
\item The seed data provides a contribution of
$\int_0^1f(x_0)dx_0$ (S). \item Each boundary data provides a
contribution of $\int_{x_0}^1 f(x)dx$ (B). \item An inner data
provides a contribution $\int_{x_j}^{x_0} \frac{f(x)dx}{1-F(x)}$
(I).
\end{itemize}
These diagrammatic-like rules allow us to schematize in a formal
way the probability associated to each configuration. For instance
in the case $k=2$, $P(k)$ has a single contribution $p_0$
represented by the formal diagram B-S-B, while for $k=3$,
$P(k)=p_0+p_1$ where $p_0$'s diagram is B-S-I-B and $p_1$'s is
B-I-S-B. It seems quite straightforward to derive a general
expression for $P(k)$, just by applying the preceding rules for
the contribution of each $C_i$. However, there is still a subtle
point to address that will become evident for the case
$P(k=4)=p_0+p_1+p_2$ (see figure \ref{k4}). While in this case
$C_1$ leads to essentially the same expression as for both
configurations in $k=3$ (and in this sense one only needs to apply
the preceding rules to derive $p_1$), $C_0$ and $C_2$ are
geometrically different configurations. These latter ones are
configurations formed by a seed, two bounding and two
\emph{concatenated} inner data, and concatenated data lead to
concatenated integrals. For instance, applying the same formalism
as for $k=3$, one come to the conclusion that for $k=4$,
\begin{equation}
p_0=\int_0^1 f(x_0)dx_0  \int_0^{x_0} \frac{f(x_1)dx_1}{1-F(x_1)}
\int_{x_1}^{x_0} \frac{f(x_2)dx_2}{1-F(x_2)} \int_{x_0}^1
f(x_3)dx_3 \int_{x_0}^1 f(x_{-1})dx_{-1}.
\end{equation}
While for the case $k=3$ every integral only depended on $x_0$
(and consequently we could integrate independently every term
until reaching the dependence on $x_0$), having two concatenated
inner data on this configuration generates a dependence on the
integrals and hence on the probabilities. For this reason, each
configuration is not equiprobable in the general case, and thus
will not provide the same contribution to the probability $P(k)$
($k=3$ was an exception for symmetry reasons). In order to weight
appropriately the effect of these concatenated contributions, we
can make use of the definition of $p_i$. Since $P(k)$ is formed by
$k-1$ contributions labelled $C_0, C_1...C_{k-2}$ where the index
denotes the number of inner data present at the left-hand side of
the seed, we deduce that in general the $k-2$ inner data have the
following effective contribution to $P(k)$:
\begin{itemize}
\item $p_0$ has $k-2$ concatenated integrals (right-hand side of
the seed). \item $p_1$ has $k-3$ concatenated integrals
(right-hand side of the seed) and an independent inner data
contribution (left-hand side of the seed). \item $p_2$ has $k-4$
concatenated integrals (right-hand side of the seed) and another 2
concatenated integrals (left-hand side of the seed). \item ...
\item $p_{k-2}$ has $k-2$ concatenated integrals (left-hand side
of the seed).
\end{itemize}
Observe that $p_i$ is symmetric with respect to the seed.\\

\noindent Including this modification in the diagrammatic rules,
we are now ready to calculate a general expression for $P(k)$.
Formally,
\begin{equation}
P(k)=\sum_{j=0}^{k-2}[S][B]^2[I]_j[I]_{k-2-j} \label{formal},
\end{equation}
where the sum extends to each of the $k-1$ configurations, the
superindex denotes exponentiation and the subindex denotes
concatenation (this latter expression can be straightforwardly
proved by induction). In order to solve it, one needs to firstly
calculate the concatenation of $n$ inner data integrals
$[I]_n\equiv I(n)$, that is
\begin{equation}
I(n)=\int_0^{x_0}
\frac{f(x_1)dx_1}{1-F(x_1)}\prod_{j=1}^{n-1}\int_{x_j}^{x_0}\frac{f(x_{j+1})dx_{j+1}}{1-F(x_{j+1})}.\label{recurrencia}
\end{equation}
The calculation of $I(n)$ is easy but quite tedious. One proceeds
to integrate equation \ref{recurrencia} step by step (first $n=1$,
then $n=2$, and so on), and a recurrence quickly becomes evident.
One can easily prove by induction that
\begin{equation}
I(n)=\frac{(-1)^{n}}{n!}\bigg[\ln
\big(1-F(x_0)\big)\bigg]^n.\label{In}
\end{equation}
According to the formal solution \ref{formal} and to equation
\ref{In}, we finally have
\begin{eqnarray}
P(k)&=&\sum_{j=0}^{k-2}\frac{(-1)^{k-2}}{j!(k-2-j)!}\int_0^1f(x_0)[1-F(x_0)]^2[\ln(1-F(x_0))]^{k-2}dx_0\nonumber
\\
&=& 3^{1-k}\sum_{j=0}^{k-2}\frac{(k-2)!}{j!(k-2-j)!} =
\frac{1}{3}\bigg(\frac{2}{3}\bigg)^{k-2}\label{teorico}
\end{eqnarray}
\noindent Surprisingly, we can conclude that for every probability
distribution $f(x)$, the degree distribution $P(k)$ of the
associated horizontal visibility graph
has the same exponential form.\\
\begin{figure}[h]
\centering
\includegraphics[width=0.50\textwidth]{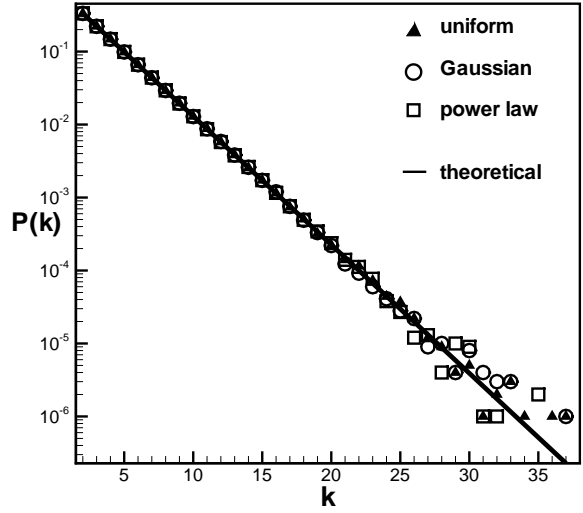}
\caption{Semi-log plot of the degree distribution of an horizontal
visibility graph associated to random series of $10^6$ data
extracted from a uniform distribution $f(x)=U[0,1]$ (triangles), a
Gaussian distribution (circles), and power law distribution
$f(x)\sim x^{-2}$ (squares). Solid line corresponds to equation
14.} \label{conectividad}
\end{figure}

\noindent In order to check further the accuracy of our analytical
results for the case of \emph{finite} time series, we have
performed several numerical simulations. We have generated random
series of $10^6$ data from different distributions $f(x)$ and have
generated their associated horizontal visibility graphs. In figure
\ref{conectividad} we have plotted the degree distribution of the
resulting graphs (triangles correspond to a series extracted from
a uniform distribution, while circles and squares correspond to
one extracted from a Gaussian and a power law distribution
$f(x)\sim x^{-2}$ respectively). The solid line corresponds to the
theoretical equation \ref{teorico}, showing a perfect agreement
with the numerics.

\section{Some other topological properties of the visibility graph}
\subsection{Degree versus height}
\begin{figure}[h]
\centering
\includegraphics[width=0.50\textwidth]{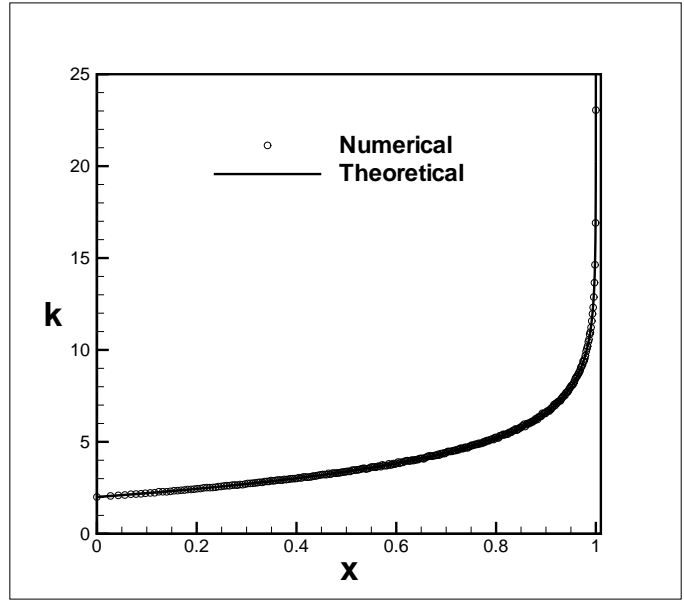}
\caption{Average degree of a node, as a function of the associated
datum's height: (circles) numeric results from a random series of
$10^6$ data extracted from a uniform distribution $f(x)=U[0,1]$.
The solid line corresponds to the theoretical prediction
eq.\ref{K_extreme}, showing a perfect agreement. It comes evident
that the hubs stand for the nodes associated to the data with
larger values (extreme events).} \label{degreeheight}
\end{figure}
An interesting aspect worth exploring is the relation between data
height and the node degree, that is, to study whether a functional
relation between the height of a datum and the degree of its
associated node holds. In this sense, let us define $P(k|x)$ as
the conditional probability that a given node has degree $k$
provided that it has height $x$. Observe that $P(k|x)$ can be
easily deduced from eq. \ref{teorico}, such that
\begin{equation}
P(k|x)=\sum_{j=0}^{k-2}\frac{(-1)^{k-2}}{j!(k-2-j)!}[1-F(x)]^2\cdot[\ln(1-F(x))]^{k-2}.
\end{equation}
Notice that probabilities are well normalized and that
$\sum_{k=2}^\infty P(k|x)=1$, independently of $x$. Now, we can
define an average value of the degree of a node associated to a
datum of height $x$, $K(x)$, in the following way
\begin{equation}
K(x)=\sum_{k-2}^\infty kP(k|x)=2-2\ln(1-F(x)).\label{K_extreme}
\end{equation}
Since $F(x)\in[0,1]$ and $\ln(x)$ are monotonically increasing
functions, $K(x)$ will also be monotonically increasing. We can
thus conclude that graph hubs (that is, the most connected nodes)
are the data with largest values, that is,
the extreme events of the series.\\
In order to check the accuracy of the theoretical prediction
within finite series, in fig. \ref{degreeheight} we have plotted
(circles) the numerical values of $K(x)$ within a random series of
$10^6$ data extracted from a uniform distribution with $F(x)=x$.
The line corresponds to eq. \ref{K_extreme}, showing a perfect
agreement.

\subsection{Local clustering coefficient distribution}
\begin{figure}[t]
\centering
\includegraphics[width=0.50\textwidth]{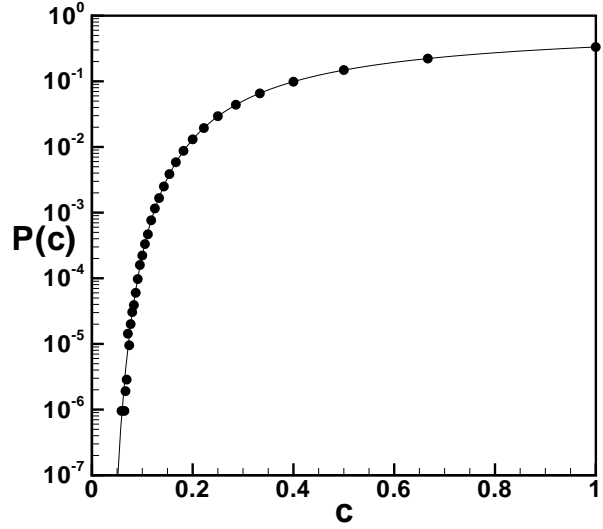}
\caption{Semi-log plot of the clustering distribution of an
horizontal visibility graph associated to random series of $10^6$
data extracted from a uniform distribution $f(x)=U[0,1]$ (dots).
The solid line corresponds to the theoretical prediction
$P(C)=(1/3)(2/3)^{2/C-2}$. In order to avoid border effects we
have imposed periodic boundary conditions in the data series.}
\label{distclust}
\end{figure}

The local clustering coefficient $C$ \cite{redes2, redes3, redes4,
redes5, redes1} of an horizontal visibility graph associated to a
random series can be easy deduced by means of geometrical
arguments. For a given node $i$, $C$ denotes the rate of nodes
connected to $i$ that are connected between each other (observe
that in this section, $C$ denotes clustering: do not mistake this
with the 'C' (configuration) of section III). In other words, we
have to calculate from a given node $i$ how many nodes from those
visible to $i$ have mutual visibility (triangles), normalized with
the set of possible triangles ${k \choose 2}$. In a first step, if
a generic node $i$ has degree $k=2$, these nodes are
straightforwardly two bounding data, hence having mutual
visibility. Thus, in this situation there exists $1$ triangle and
$C(k=2)=1$. Now if a generic node $i$ has degree $k=3$, one of its
neighbors will be an inner data, which will only have visibility
of one of the bounding data (by construction). We conclude that in
this situation we can only form $2$ triangles out of $6$ possible
ones, thereby $C(k=3)=2/6$. In general, for a degree $k$ we can
form $k-1$ triangles out of ${k \choose 2}$ possibilities, and
then:
\begin{equation}
C(k)=\frac{k-1}{{k \choose 2}}=\frac{2}{k},
\end{equation}
what indicates a so called hierarchical structure
\cite{hierarchical}. This relation between $k$ and $C$ allows us
to deduce the local clustering coefficient distribution $P(C)$:
\begin{eqnarray}
&& P(k)=\frac{1}{3}\bigg(\frac{2}{3}\bigg)^{k-2} = P(2/C) \nonumber\\
&&P(C)=\frac{1}{3}\bigg(\frac{2}{3}\bigg)^{2/C-2}\label{cluster}
\end{eqnarray}
To check the validity of this latter relation within finite
series, in figure \ref{distclust} we depict the clustering
distribution of an horizontal visibility graph associated to a
random series of $10^6$ data (dots) obtained numerically. The
solid line corresponds to the theoretical prediction (equation
\ref{cluster}), in excellent agreement with the
numerics.\\
\subsection{Long distance visibility, mean degree and mean path length}
\noindent In order to derive the scaling of the mean path length
\cite{redes1}, let us first calculate the probability $P(n)$ that
two data separated by $n$ intermediate data be two connected nodes
in the graph. Consider again a time series extracted from a random
variable $X$ with probability distribution $f(x)$ and $x \in
[0,1]$, and let us construct its associated horizontal visibility
graph. An arbitrary value $x_0$ from this series will `see' $x_n$
(and consequently will be connected to node $x_n$ in the graph)
iff $x_i < \min(x_0, x_n)$ for all $x_i, \  i=1,2,...,n-1$. Then
$P(n)$ can be expressed as:
\begin{eqnarray}
&&P(n)=\int_{0}^{1}\int_{0}^{1}f(x_0)f(x_n)dx_0dx_n\int_{0}^{\min(x_0,
x_n)}\dots\nonumber\\
&&\dots\int_{0}^{\min(x_0, x_n)}f(x_1)\ldots f(x_{n-1})dx_1\dots
dx_{n-1}
\end{eqnarray}
Since the integration limits are independent, rewriting
$x\equiv\min(x_0,x_n)$ we have
\begin{equation}
P(n)=\int_0^1\int_0^1f(x_0)f(x_n)F^{n-1}(x)dx_0dx_n.\label{pnf2}
\end{equation}
We can fix $x_0$ and move $x_n$ without loss of generality, such
that the latter equation can be expressed as
\begin{equation}
P(n)=\underbrace{\int_{0}^{1}\int_{0}^{x_0}f(x_0)f(x_n)F^{n-1}(x_n)dx_0dx_n}_{\small{\textrm{the
minimum here is
$x_n$}}}+\underbrace{\int_{0}^{1}\int_{x_0}^{1}f(x_0)f(x_n)F^{n-1}(x_0)dx_0dx_n}_{\small{\textrm{the
minimum here is $x_0$}}}
\end{equation}
Applying the definition of $F(x)$ and the relation \ref{f-F}, with
a little calculus we get
\begin{eqnarray}
P(n)&=&\bigg(\frac{1}{n}-1\bigg)\int_0^1f(x_0)F^n(x_0)dx_0+\int_0^1f(x_0)F^{n-1}(x_0)dx_0\nonumber
\\
&=&\frac{2}{n(n+1)}.
\end{eqnarray}
Observe that $P(n)$ is again independent of the probability
distribution of the random variable $X$. Notice that the latter
result can also be obtained, alternatively, with a purely
combinatorial argument that reads as it follows. Take a random
series with $n+1$ data and choose its two largest values. This
latter pair can be placed with equiprobability in $n(n+1)$
positions, while only two of them are such that the largest values
are placed at distance $n$, so we get
$P(n)=\frac{2}{n(n+1)}$ on agreement with the previous development.\\

\noindent At this point, we can calculate the mean degree $<k>$ of
the horizontal visibility graph:
\begin{equation}
<k>=\sum
kP(k)=\sum_{k=2}^{\infty}\frac{k}{3}\Bigg{(}{\frac{2}{3}}\Bigg{)}^{k-2}
= 4, \label{media_total}
\end{equation}
\noindent that we can recover from $P(n)$ as:
\begin{equation}
<k> = 2\sum_{n=1}^{\infty}P(n)=4. \label{media_total_1}
\end{equation}

\begin{figure}[t]
\centering
\includegraphics[width=0.50\textwidth]{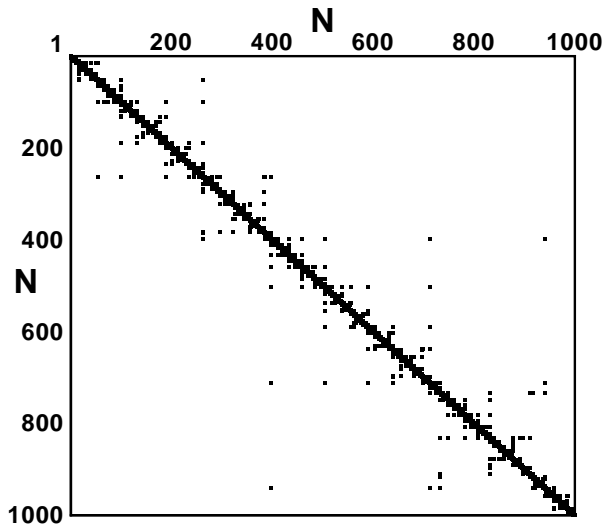}
\caption{Adjacency matrix of an horizontal visibility graph
associated to a random series of $10^3$ data. \label{adja}}
\end{figure}

\begin{figure}[t]
\centering
\includegraphics[width=0.50\textwidth]{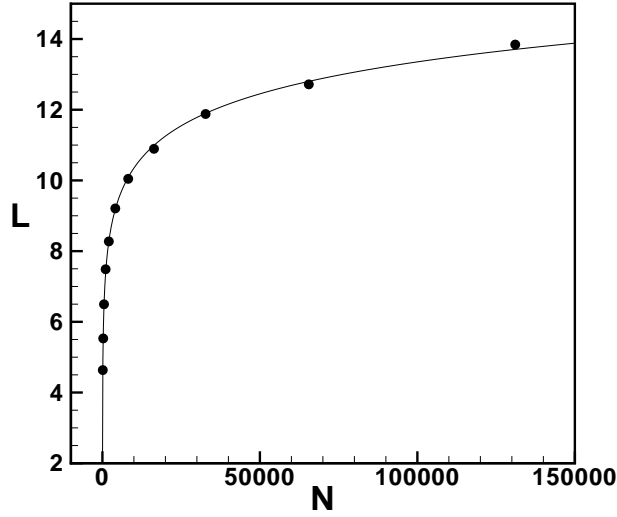}
\caption{Mean path length $L(N)$ of an horizontal visibility graph
associated to random series of $N=2^7, 2^8, \dots 2^{17}$ data
(dots). The solid line corresponds to the better fit $L(N)=
1.3\log(N)-1.7$.\label{dist_media}}
\end{figure}

\noindent Now, for illustration purposes, in figure \ref{adja} we
show the adjacency matrix \cite{redes1} of the horizontal
visibility graph associated to a random series of $10^3$ data (the
entry $i,j$ is filled in black if nodes $i$ and $j$ are connected,
and left blank otherwise). Since every data $x_i$ has visibility
of its first neighbors $x_{i-1}$, $x_{i+1}$, every node $i$ will
be connected by construction to nodes $i-1$ and $i+1$: the graph
is thus connected. Observe in figure \ref{adja} that the graph
evidences a typical homogeneous structure: the adjacency matrix is
predominantly filled around the main diagonal. Furthermore, the
matrix evidences a superposed sparse structure, reminiscent of the
visibility probability $P(n)=2/(n(n+1))$ that introduces some
shortcuts in the horizontal visibility graph, much in the vein of
the Small-World model \cite{redes2.0}. Here the probability of
having these shortcuts is given by $P(n)$. Statistically speaking,
we can interpret the graph's structure as quasi-homogeneous, where
the size of the local neighborhood increases with the graph's
size. Accordingly, we can approximate its mean path length $L(N)$
as:
\begin{equation}
L(N)\approx\sum_{n=1}^{N-1} nP(n)=\sum_{n=1}^{N-1} \frac{2}{n+1} =
2\log(N)+2(\gamma-1)+O(1/N), \label{L}
\end{equation}
where we have made use of the asymptotic expansion of the harmonic
numbers and $\gamma$ is the Euler-Mascheroni constant. As can be
seen, the scaling is logarithmic, denoting that the horizontal
visibility graph associated to a generic random series is
Small-World \cite{redes2.0}, according to what figure 6 suggested.
In figure \ref{dist_media} we have plotted the numerical results
of $L(N)$ (dots) of an horizontal visibility graph associated to
several random series of increasing size $N=2^7, 2^8, \dots
2^{17}$. The solid line corresponds to the best fit $L(N)=
1.3\log(N)-1.7$.

\section{Application of the theory to discriminate chaotic series}
\begin{figure}[t]
\centering
\includegraphics[width=0.50\textwidth]{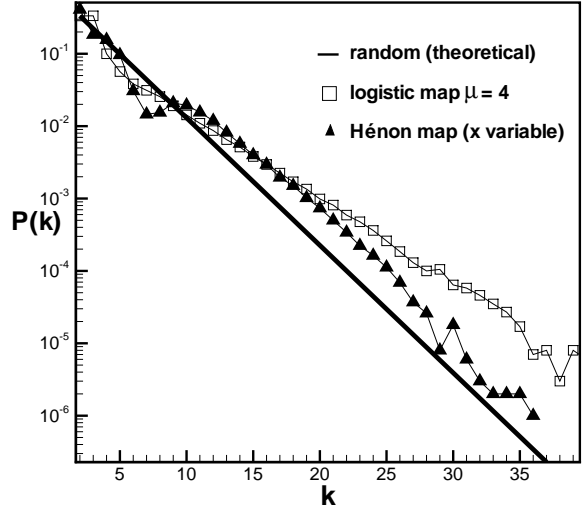}
\caption{Semi-log plot of the degree distribution of several
horizontal visibility graphs associated to: (solid line)
theoretical prediction for random series (equation \ref{teorico}),
(squares) time series of $10^6$ extracted from a logistic map
$x_{t+1}=\mu x_t(1-x_t)$ in the fully chaotic region $\mu=4$,
(black triangles) time series of $10^6$ extracted from $x$
variable of the H\'{e}non map $(x_{t+1},y_{t+1})=(y_t+1-ax_t^2,
bx_t)$ in the fully chaotic region ($a=1.4$, $b=0.3$).}
\label{comparison}
\end{figure}
So far we have presented exact results on the topological
properties of graphs associated to series of i.i.d. random
variables (random series from now on) via the horizontal
visibility algorithm. The very first application of this theory
can be found in the task of discriminating a random signal from a
chaotic one. The task of identifying random processes and more
concretely discriminating (low dimensional) deterministic chaotic
systems from stochastic processes has been extensively studied in
the last decades (see for instance \cite{procaccia1, Farmer,
sugihara, tsonis, kaplan, PRL2007, libro_TSA}). Essentially, all
methods that have been introduced so far rely on two major points:
Firstly, chaotic systems have a finite dimensional attractor,
whereas stochastic processes arise from an infinitely dimensional
one. Being able to reconstruct this latter attractor is thus a
clear evidence showing that the time series has been generated by
a deterministic system. Secondly, deterministic systems evidence,
as opposed to random ones, short-time prediction: the difference
between the time evolution of two nearby states will remain rather
low for regular systems and increase exponentially fast for
chaotic ones, while for stochastic processes this difference
should be randomly distributed. Whereas several algorithms relying
on the preceding concepts are nowadays available, the great
majority of them are purely numerical and/or usually complicated
to perform, computationally speaking (these difficulties are
eventually more acute for noisy series \cite{noise} or high
dimensional chaotic ones \cite{spatiotemporal}). Furthermore, even
the discrimination between a chaotic series and a series of i.i.d.
random variables, something that an autocorrelation function or a
power spectra fails to do but some other methods such as
recurrence plots can \cite{kaplan2} is nontrivial when the chaotic
degree of the series is high, or even when such series is polluted
with noise. All these complications provide motivation for a
search for new methods that can directly distinguish, in a
reliable way, random from chaotic time series, prior to
quantifying the dimension \cite{kennel} and without needs for
additional sophisticated techniques such as surrogate data
\cite{surrogate} or noise reduction methods \cite{noise}. In the
preceding sections we have proved that the horizontal visibility
graph associated to a random series has well-defined and universal
degree distribution, local clustering distribution and $P(n)$,
independent of the shape of the random probability distribution
$f(x)$. These theorems guarantee that horizontal visibility graphs
with other topological properties are not uncorrelated random
series. In what follows we explore the reliability of the method
to distinguish uncorrelated randomness from chaos in finite
series.

\subsection{Low-dimensional chaos}
In order to test the practical usefulness of this method, we have
generated the horizontal visibility graph of several noise-free
chaotic series, and have calculated numerically their degree
distribution. We have restricted our analysis to discrete systems
(maps), but the method is also extensible to flows (in that case
the null hypothesis would no longer be white noise but Brownian
motion \cite{hurst}). In figure \ref{comparison} we have plotted
in semi-log the results of these simulations. Compare it with
figure 3. In every case and by simple visual inspection we can
conclude that $P(k)$ deviates from equation \ref{teorico}: the
method is able to easily distinguish randomness from
low-dimensional chaos (similar results are obtained with $P(n)$
and $P(C)$, but $P(k)$ works better as
discriminator).\\

Observe at this point that if we shuffle the series data and
reproduce the analysis, we would find a degree distribution that
now would correspond to equation \ref{teorico}, since shuffling
breaks the temporal correlations of the series: such shuffled
series would be equivalent to a random series extracted from a
probability distribution equal to the system's probability measure
(the beta distribution in the case of the Logistic map). We can
deduce that the algorithm captures temporal correlations of time
series, and that $P(k)$ plays the role of an autocorrelation
function, but with the additional ability of capturing nonlinear
correlations. Observe also that this method neither works on the
time nor on the frequency domain, since it only makes use
of topological features.\\
\subsection{Noisy chaotic series}
\begin{figure}[h]
\centering
\includegraphics[width=0.50\textwidth]{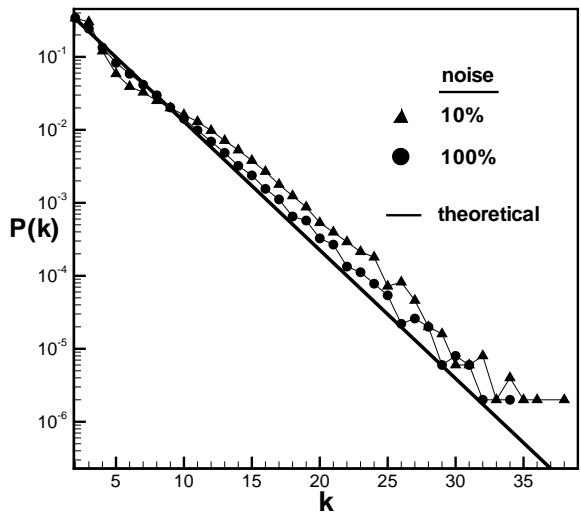}
\caption{Semi-log plot of the degree distribution of an horizontal
visibility graph associated to:  (triangles) noisy chaotic series
of $10^5$ data extracted from the Logistic map ($\mu=4$) with a
measurement noise level of $10\%$ (by amplitude), (circles) idem
but for noise level of $100\%$. The solid line corresponds to the
theoretical prediction for random series $P(k)=(1/3)(2/3)^{k-2}$.}
\label{noise_series}
\end{figure}

It is well known that standard methods evidence problems when
noise is present in chaotic signals, since even a small amount of
noise can destroy the fractal structure of a chaotic attractor and
mislead the calculation of chaos indicators such as the
correlation dimension or the Lyapunov exponents \cite{noise}. In
order to check the algorithm's robustness, we have introduced an
amount of white noise (measurement noise) in a signal extracted
from a fully chaotic Logistic map ($\mu=4.0$). In figure
\ref{noise_series} we plot the degree distribution of its
associated visibility graph. Remarkably, the algorithm still
discriminates noisy chaotic behavior from randomness even when the
noise level reaches the $100\%$ of the signal amplitude.

\subsection{High dimensional chaos: Coupled Map Lattice}
\begin{figure}[h]
\includegraphics[width=0.50\textwidth]{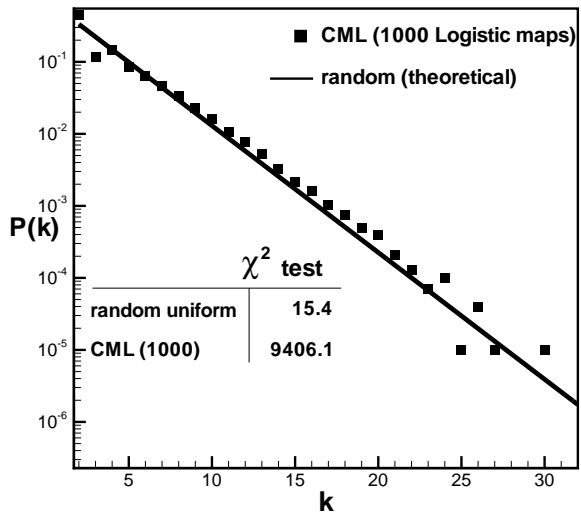}
\caption{Semi-log plot of the degree distribution of an horizontal
visibility graph associated to:  (squares) chaotic series of
$10^5$ data extracted from a CML of 1000 Logistic maps with
$\mu=4$ and a coupling strength $\epsilon=0.4$ (see the text). The
solid line corresponds to the theoretical prediction for random
series $P(k)=(1/3)(2/3)^{k-2}$. Inner box: values of $\chi^2$
goodness of fit test between eq. \ref{teorico} and (i) a random
series extracted from a uniform distribution (reference value),
(ii) the CML series. While visually speaking the deviations from
eq.\ref{teorico} are not here as evident as in the case of low
dimensional chaotic series, a $\chi^2$ test clearly rejects the
hypothesis of randomness (the critical values for $10\%$, $5\%$
and $1\%$ level of significance are 35.2, 38.1 and 41.7
respectively, very far from the value of the test statistic
9406.1): the algorithm distinguishes high-dimensional chaos from
randomness.} \label{CML}
\end{figure}

Standard methods of phase space reconstruction also demonstrate
computational problems when the attractor dimension is large
\cite{spatiotemporal}. Here we have generated high dimensional
chaotic series by means of the so called Coupled Map Lattice (CML)
\cite{libro_manrubia}, a paradigmatic formalism for spatiotemporal
chaos, widely used to model chaotic extended systems including
fully-developed turbulence and pattern formation problems. We have
coupled $N=1000$ Logistic maps $x_t^i,\ i=1,...,N$ such that
$x_{t+1}^i = f(\epsilon/2x_t^{i-1} + (1-\epsilon)x_t^i +
\epsilon/2x_t^{i+1})$, where $f(x)=4x(1-x)$ and $\epsilon=0.4$ is
the coupling strength (note that such a system exhibits high
dimensional chaos with an estimated attractor dimension $D_2=800$
\cite{spatiotemporal}). In figure \ref{CML} we have plotted the
degree distribution of its associated visibility graph along with
the theoretical prediction for a random series. While the
deviations from eq. \ref{teorico} are not as evident as for low
dimensional chaotic series, a $\chi^2$ clearly rejects the
hypothesis of randomness:
the method distinguishes randomness from high dimensional chaos. \\

\subsection{Topological properties of chaotic series}
Observe in fig. \ref{comparison} that the series extracted from
the Logistic and H\'{e}non maps seem to have an associated
visibility graph with a degree distribution which has an
exponential tail, yet different to eq. \ref{teorico}. This
characteristic can be explained as follows: First, the tail of
$P(k)$ is related to the hubs degree. Hubs correspond to the data
series that have largest visibility. These are, according to eq.
\ref{K_extreme}, extreme events in the series, whose degree is
truncated by other extreme data (statistically speaking).
Accordingly, the tail of $P(k)$ essentially reduces to calculate
the probability distribution of recurrence times in the series.
Within random series, notice that this distribution is
straightforwardly exponential (recurrence times in a Poisson
process are exponentially distributed \cite{Feller}), consistent
with eq.\ref{teorico}. Within chaotic series, recurrence time
statistics are related \cite{PRT5} to the concept Poincar\'{e}
recurrence time (PRT) \cite{zaslavsky}, which measures the time
interval between two consecutive visits of a trajectory to a
finite size region of the phase space. As a matter of fact, it has
been shown that Poincar\'{e} recurrence times are exponentially
distributed in several hyperbolic chaotic systems, including the
Logistic and H\'{enon} maps (see \cite{carleson} and references
therein). We conjecture that the functional form of $P(k)$ is
closely related for chaotic series with their associated
Poincar\'{e} recurrence time distribution (which deviate from the
Poissonian statistics (eq. \ref{teorico}) due to deterministic
effects), something that will be addressed in future work.

\subsection{Stochastic processes versus chaos}
The task of distinguishing determinism from a generic stochastic
process (e.g. fractional Brownian motion, high order Markov
models, etc) is more general and goes well beyond the scope of
this work, since our theory only addresses series of i.i.d. random
variables (uncorrelated random series). However, in a recent work
\cite{hurst} it has been shown that fractional Brownian motions
and colored noise series map into scale free visibility graphs,
which clearly differ from the functional form of $P(k)$ for
chaotic series and from i.i.d. theory. In this sense we conjecture
that the visibility algorithm efficiently discriminates not only
uncorrelated randomness from chaos but also more complicated
stochastic processes such as colored noise or fractional Brownian
motion.

\section{Some conjectured random like series: decimal expansion of normal numbers}
A real number $R$ (which can be understood as a series if we pick
its decimal expansion) is defined as a normal number if for all
integer $k$, any given $k-$tuple is equally likely in the
$k-$expansion of $R$; that is to say, the digits of a real number
show a uniform distribution in every base \cite{normal}. For
instance, in a decimal base, if number $R$ is normal then every
string of size $k$ is equally likely to appear: for $k=4$ the
string $3254$ is as likely as $1234$, and this holds for all $k$.
It is a well-known result from measure theory that a real number
chosen at random is absolutely normal with probability 1.
Interestingly, many fundamental constants such as $\pi, e$ or
common irrational numbers such as $\ln 2$ or $\sqrt{2}$ are
conjectured to be normal, but not a single proof exists so far
\cite{normal}. Now, the degree distribution of a visibility graph
associated to the series generated by the decimal expansion of a
normal number should follow equation \ref{teorico}. In other
words, a deviation from eq. \ref{teorico} would imply the
non-normality of a given number. In table \ref{table1} we have
reported the values of a $\chi^2$ goodness-of-fit test between the
degree distribution of graphs associated to the decimal expansion
of several conjectured normal numbers (series of $N=100000$ data)
and equation \ref{teorico}. The same test has been performed for
the case of a random series extracted from a uniform distribution
of the same size, for the sake of comparison. As expected, the
null hypothesis of normality cannot be rejected. Note that this
procedure can easily extend to other number theoretic sequences
which are also
conjectured to be random.\\

\begin{table}
\begin{ruledtabular}
\begin{tabular}{ccddd}
Series&$\chi^2$\\
\hline
Decimal expansion of $\pi$&19.9\\
Decimal expansion of $e$&20.2\\
Decimal expansion of $\ln 2$&22.34\\
Random series extracted from uniform distribution &23.1\\
\end{tabular}
\end{ruledtabular}
\caption{\label{table1}$\chi^2$ goodness-of-fit test between the
degree distribution of visibility graphs associated to several
number theoretical sequences and the theoretical prediction for
random series. The number theoretical sequences are constructed
from the decimal expansion of the first $6\cdot10^5$ digits of
$\pi, e$ and $\ln 2$, grouped by tuples of 6 elements, what
provides series of $10^5$ data respectively. As expected, the
normality of $\pi, e$ and $\ln 2$ is not rejected by the $\chi^2$
test.}
\end{table}

\subsection{Note on flows}
Notice that the theory that we have developed in sections II to IV
addresses a series of i.i.d. variables, that is, a discrete
series. Accordingly, we have compared the results obatained from
chaotic maps or from the decimal expansion of numbers to this
i.i.d. theory, that is, discrete data. Now, it is not
straightforward to compare this theory with visibility graphs
extracted from flows (continuous series), since in any
discretization of a flow some continuity properties are present,
something that is not assumed a priori in the i.i.d. theory. This
will be addressed in further work.

\section{Concluding remarks}
In this work we have introduced the horizontal visibility
algorithm, an algorithm that maps time series into graphs which is
inspired in the so called visibility algorithm
\cite{visibilidad_pnas}. The present algorithm is quite similar to
the latter, yet analytically solvable. Accordingly, we have
obtained exact results on several properties of the horizontal
visibility graph associated to generic uncorrelated random series,
and numerical simulations confirmed its reliability for finite
series. Concretely, the degree distribution of the graph has an
exponential form $P(k)=(1/3)(2/3)^{k-2}$, the clustering
coefficient $C$ has a probability distribution
$P(C)=(1/3)(2/3)^{2/C-2}$ and the mean path length scales with the
system's size in a logarithmic fashion, evidencing the Small-World
phenomenon \cite{redes2.0}. Since the results are independent of
the distribution from which the series was generated, we conclude
that every uncorrelated random series must have the same
horizontal visibility graph, and in particular the same degree
distribution. Thereby, this algorithm can be used as a simple test
for discriminating uncorrelated randomness from chaos. Concretely,
we have shown that the method can perfectly distinguish between
random series (different probability distributions) that indeed
follow the theoretical prediction and chaotic series (logistic
map, tent map, H\'{e}non map) that clearly deviate from the
theory. This extends to chaotic series polluted with noise and
even to high dimensional chaotic series (coupled map lattice).

\noindent Observe that this method diverges from the standard
algorithms introduced so far \cite{libro_TSA} since it makes use
of graph theoretical techniques to characterize nonlinear temporal
correlations of the series, and its recipe is straightforward: (i)
construct a visibility graph from the series under study, (ii)
compute its degree distribution $P(k)$ and compare it to eq.
\ref{teorico}. A visual inspection (or eventually a $\chi^2$
goodness-of-fit test between $P(k)$ and equation \ref{teorico} if
needed) allows us to reject the hypothesis that the series is
random. The algorithm is direct and has low computational cost, as
opposed to several standard methods. Furthermore, it is not just
empirical since it is based in exact results. It is worth
emphasizing that its purpose is not to quantify chaos but to
easily discriminate chaos from uncorrelated randomness. For
practical purposes, the method should be used as a reliable
preliminary test when looking for deterministic fingerprints in
time series (in this sense, once we have checked that $P(k)$ has
an exponential tail that deviates from equation \ref{teorico},
embedding methods should be applied to the series). Whether this
algorithm is also able to quantify chaos, as well as the relation
between standard chaos indicators (Lyapunov exponents, correlation
dimension, etc) and the topological
properties of the visibility graphs are open problems for further research.\\
It is also worth commenting that in a preceding work it has been
shown that the visibility algorithm is also able to identify
colored noise series ($f^{-\beta}$ noises, fractional Brownian
motion), since their associated visibility graphs are scale-free
\cite{hurst}, and an algebraic relation between the exponent of
the power law degree distribution and the Hurst exponent of the
time series exists. In this sense, the visibility algorithm can
also discriminate chaos from colored noise.

\section{acknowledgments}
The authors acknowledge the interesting comments of two anonymous
referees. This work is partially supported by the spanish ministry
of science under grant FISXXXXXX.

\end{document}